\documentclass[10pt,superscriptaddress,tightenlines,twocolumn,amsmath,amssymb,aps, pra]{revtex4-2}
\UseRawInputEncoding
\usepackage{graphicx}
\usepackage{dcolumn}
\usepackage{bm}
\usepackage{color}
\usepackage{amsmath}
\usepackage{amssymb}
\usepackage{subfigure}
\usepackage{braket}
\usepackage{indentfirst}
\usepackage{mathrsfs}
\usepackage{booktabs}
\usepackage{amssymb}
\usepackage{amsmath}
\usepackage{amsfonts}
\usepackage{graphicx}
\usepackage{makecell}
\usepackage{hyperref}
\usepackage{makecell}
\usepackage{color}%
\usepackage{lmodern}
\usepackage{lineno}
\usepackage{bbding}
\usepackage{xcolor}
\usepackage{enumitem}
\usepackage{multirow}

\colorlet{RED}{red}
\colorlet{BLACK}{black}

\providecommand{\U}[1]{\protect\rule{.1in}{.1in}}

\hypersetup{colorlinks=true, linkcolor=red, citecolor=blue, urlcolor=black}

\def\bra#1{\mathinner{\langle{#1}|}}
\def\ket#1{\mathinner{|{#1}\rangle}}

\def\fid#1#2{\left\langle{#1}|{#2}\right\rangle}

\begin{document}


\title{Experimental Coherent One-Way Quantum Key Distribution with Simplicity and Practical Security}


\author{Xiao-Yu Cao}\thanks{These authors contributed equally to this work.}
\author{Xiao-Ran Sun}\thanks{These authors contributed equally to this work.}
\author{Ming-Yang Li}
\author{Yu-Shuo Lu}
\affiliation{National Laboratory of Solid State Microstructures and School of Physics, Collaborative Innovation Center of Advanced Microstructures, Nanjing University, Nanjing 210093, China}
\affiliation{School of Physics and Key Laboratory of Quantum State Construction and Manipulation (Ministry of Education), Renmin University of China, Beijing 100872, China}
\author{Hua-Lei Yin}\email{hlyin@ruc.edu.cn}
\affiliation{School of Physics and Key Laboratory of Quantum State Construction and Manipulation (Ministry of Education), Renmin University of China, Beijing 100872, China}
\affiliation{National Laboratory of Solid State Microstructures and School of Physics, Collaborative Innovation Center of Advanced Microstructures, Nanjing University, Nanjing 210093, China}
\affiliation{Beijing Academy of Quantum Information Sciences, Beijing 100193, China}
\author{Zeng-Bing Chen}\email{zbchen@nju.edu.cn}
\affiliation{National Laboratory of Solid State Microstructures and School of Physics, Collaborative Innovation Center of Advanced Microstructures, Nanjing University, Nanjing 210093, China}

\begin{abstract}
Coherent one-way quantum key distribution (COW-QKD) has been widely investigated, and even been deployed in real-world quantum network. However, the proposal of the zero-error attack has critically undermined its security guarantees, and existing experimental implementations have not yet established security against coherent attacks. In this work, we propose and experimentally demonstrate an information-theoretically secure COW-QKD protocol that can resist source side-channel attacks, with secure transmission distances up to 100 km. Our system achieves a secure key rate on the order of kilobits per second over 50 km in the finite-size regime, sufficient for real-time secure voice communication across metropolitan networks. Furthermore, we demonstrate the encrypted transmission of a logo with information-theoretic security over 100 km of optical fiber. These results confirm that COW-QKD can simultaneously provide simplicity and security, establishing it as a strong candidate for deployment in small-scale quantum networks.
\end{abstract}


\maketitle

\section{\label{introduction}Introduction}
Quantum communication aims to provide information-theoretic security for various communication tasks~\cite{wehner2018quantum,li2024asynchronous,azuma2023quantum,bozzio2024quantum} and has achieved significant progress in multiple subfields, including quantum key distribution (QKD)~\cite{bennett2014quantum,lo2012measurement,lucamarini2018overcoming,xie2022breaking,zeng2022mode,yin2016MDI404,liu2023experimental,pittaluga2025long}, quantum conference key agreement~\cite{fu2015long,proietti2021experimental,yang2024experimental,xie2024multi,du2025experimental,lu2024repeater}, quantum digital signatures~\cite{clarke2012experimental,dunjko2014quantum,yin2023experimental,du2025chip}, quantum secure direct communication~\cite{sheng2022one,zhang2022realization,ying2025passive,yang2025300}, quantum secret sharing~\cite{wang2024experimental,xiao2025experimental,zhang2024device,zhang2025device}, and so on~\cite{schiansky2023demonstration,cao2024experimental,shen2025experimental}. QKD, one of the most mature subfields, provides unconditional security for key sharing between two trusted parties~\cite{bennett2014quantum,shao2025high}. 
Early QKD protocols relied on single-photon and entangled-photon sources. Subsequently, protocols employing coherent-state sources gained widespread adoption due to their practical advantages~\cite{scarani2009security,xu2020secure}. However, the original coherent-state-based schemes were vulnerable to photon-number-splitting attacks~\cite{huttner1995quantum, brassard2000limitations}. The decoy-state method~\cite{wang2005beating,lo2005decoy} was introduced to address this attack and has since become a standard technique. Nevertheless, decoy-state protocols face the threat of side-channel attacks caused by device imperfections, which inevitably results in a substantial decrease in the secure key rate~\cite{xu2020secure,tamaki2016decoy,huang2018quantum,gnanapandithan2025hidden}.

Distributed-phase-reference scheme~\cite{inoue2002differential,stucki2005fast} is another approach to resist photon-number-splitting attacks, and coherent one-way (COW) QKD~\cite{stucki2005fast} is one of the prime protocols in this category. 
In COW-QKD, each logic bit is encoded in the arrival time of a pair of optical pulses, where each pulse is either a vacuum or a non-vacuum state. By employing a unique encoding rule, COW-QKD simplifies its experimental requirement, making it highly suitable for practical applications. Due to its simple experimental setup, COW-QKD has been widely implemented~\cite{stucki2009continuous,stucki2009high,walenta2014fast,korzh2015provably,sibson2017chip-based,sibson2017integrated,roberts2017modulator-free,de2021real}.
Additionally, considering there is no side channel in the vacuum state, Cow scheme may provide enhanced robustness against source side channels.
Since the system only requires vacuum and non-vacuum states, it can avoid most side-channel attacks~\cite{zhang2022experimental,jiang2023side}, including those arising from time domain~\cite{huang2018quantum} and hidden multidimensional modulation attacks~\cite{gnanapandithan2025hidden}.

The original upper bound on the secret key rate was established under a specific class of collective attacks~\cite{branciard_upper_2008}, which has served as the security foundation for most COW schemes over the past decade.
However, the zero-error attack proposed invalidates the previous security framework of COW-QKD~\cite{gonzalez-payo_upper_2020,trenyi_zero-error_2021}. Previous schemes have overestimated both the achievable key rate and transmission distance. 
To address zero-error attack, Ref.~\cite{lavie2022improved} introduces a variant of COW-QKD that improves key rates over long distances. But the protocol encodes each bit into a group of three pulses, which increases experimental complexity, and lacks a security proof in the finite-key regime.
More recently, a variant of COW-QKD has been proposed~\cite{gao2022simple,li2024finite}. Without introducing additional quantum state modulation, the protocol ensures security against coherent attacks under finite-key conditions.
Its security is guaranteed by estimating the upper bound on the phase error rate of the equivalent virtual entanglement protocol rather than relying on interference visibility measurements.

\begin{table*}[ht]
\centering
\caption{\textbf{Comparison between our implementation and experimental demonstrations of COW-QKD.} The focus is on the security level, the ability to resist zero-error attack, and whether finite-key analysis has been performed. \textcolor{black}{The three previous works mentioned in this table can only resist specific types of collective attacks~\cite{branciard_upper_2008}.}} 
\label{table_comparison}
\renewcommand\arraystretch{1.4}
\begin{tabular}{w{l}{3cm} w{c}{3cm} w{c}{3cm} w{c}{3cm}}
\toprule
& \makecell[c]{Security} & \makecell[c]{Zero-error attack} & \makecell[c]{Finite size}\\ 
\midrule
             This work & Coherent &  \Checkmark & \Checkmark\\
            \small Korch \emph {et~al.} \small \cite{korzh2015provably}&  Collective  & \XSolidBrush &\Checkmark\\
            \small Sibson \textit{et al.} \small \cite{sibson2017chip-based,sibson2017integrated}&  Collective  & \XSolidBrush & \XSolidBrush\\
             \small De Marco \textit{et al.} \small \cite{de2021real}&   Collective  & \XSolidBrush & \XSolidBrush \\ 
\bottomrule
\end{tabular}
\end{table*}

\begin{figure}
  \centering
  \includegraphics[width=0.9\columnwidth]{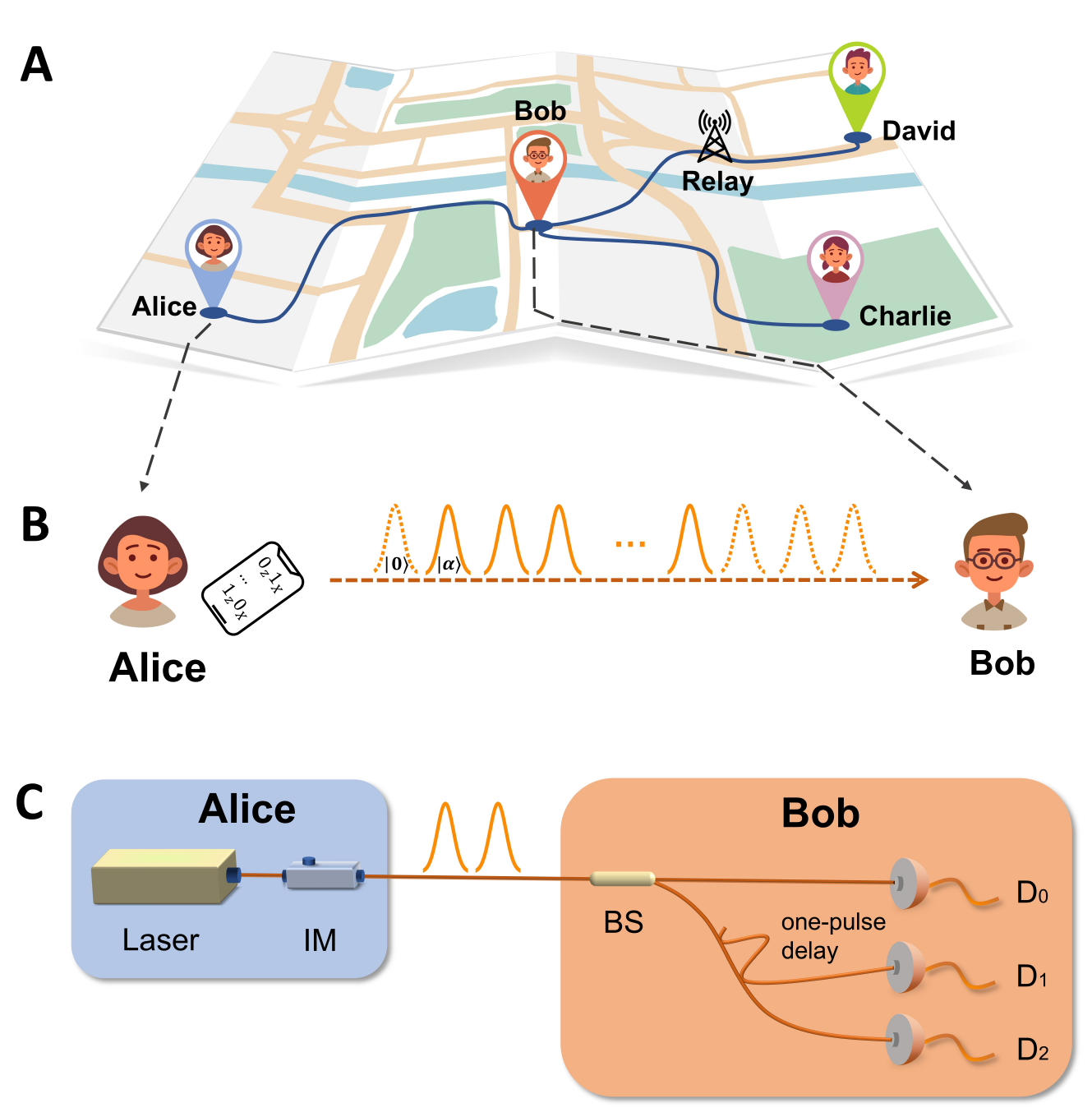}
  \caption{\textbf{Overview of the COW-QKD network architecture and experimental implementation.}
  \textbf{a.} Schematic of a quantum network. 
The network consists of senders, receivers, and intermediate relays. The sender prepares quantum states, while the receiver performs measurements. Relays, capable of both state preparation and measurement, act as intermediate nodes to extend the communication distance between users.
\textbf{b.} Conceptual demonstration of COW  protocol between two users.
Alice, acting as the sender, encodes logic bits onto optical pulses. This scheme requires only the generation of vacuum and non-vacuum states, enabling a simple modulation requirement. Bob serves as the receiver and can also function as a relay, thereby extending the achievable distance for secure key distribution. Owing to its simple experimental requirements for quantum state preparation, COW-QKD is suitable for photonic chip integration, enabling the development of portable transmitter modules.
\textbf{c.} Experimental implementation of the proposed COW-QKD.
This scheme employs the existing structure of COW protocol without any additional experimental requirement, preserving the simplicity of the implementation. IM, intensity modulation; BS, beam splitter. We gratefully acknowledge the icons designed by kalstud from www.flaticon.com, which are used in this figure.\label{cow_scheme}}
\end{figure}

In this work, we report the experimental demonstration of COW-QKD with practical security against source side-channel attacks and coherent attacks, achieving key rates over fiber links of 25, 50, 75, and 100 km. At 100 km, the secure key rate reaches 29 bps. We find that quantum state modulation in this scheme involves only vacuum and non-vacuum states, it avoids vulnerabilities to source side-channel attacks, thereby offering an advantage over traditional decoy-state schemes.
Compared to previous works (Table~\ref{table_comparison}), our implementation not only provides security against zero-error attack, but also achieves resistance to coherent attacks in the finite-key regime. 
Furthermore, a 6.13 kByte figure is encrypted with the secure keys generated over 100 km and successfully decrypted at the receiver with the shared keys. As illustrated in Fig.~\ref{cow_scheme}, the proposed COW scheme employs the same experimental setup as the original scheme~\cite{stucki2005fast}, making it fully compatible with existing quantum network infrastructures and well suited for photonic integration~\cite{sibson2017chip-based,sibson2017integrated}.  Consequently, \textcolor{black}{COW-QKD can benefit from advances in photonic quantum technologies ~\cite{wei2020high,pelucchi2022potential,li2023high,lin2025integrated,li2025surpassing}, which enable lower system cost, improved phase stability, reduced insertion losses, and higher achievable repetition rates.} Given its simplicity and compatibility, COW-QKD is a promising candidate for specific application scenarios, offering a valuable extension to future quantum network architectures.

\section{Protocol description}
Here, we give a detailed introduction to the steps of COW-QKD implemented~\cite{li2024finite}.

\textit{(\romannumeral1) Preparation}.---Alice chooses $Z$ basis and $X$ basis randomly, with probability $p_z$ and $p_x$ ($p_x = 1- p_z$), respectively. 
In the $k_{th}$ round of state preparation, if the $Z$ basis is chosen, logic bit $0$ is encoded as $\ket{0_k} = \ket{0}_{2k-1}\ket{\alpha}_{2k}$, while logic bit $1$ is encoded as $\ket{1_k} = \ket{\alpha}_{2k-1}\ket{0}_{2k}$. For the $X$ basis, Alice randomly prepares $\ket{0}_{2k-1}\ket{0}_{2k}$ with probability $p_0$, and $\ket{\alpha}_{2k-1}\ket{\alpha}_{2k}$ with probability $p_{\alpha} = p_x - p_0$. The prepared pulses are sent to Bob through optical fibers.

\textit{(\romannumeral2) Measurement}.---Bob applies a passive basis choice with a $30:70$ biased beam splitter (BS). The port with the 30\% splitting ratio is used for $Z$ basis measurement, while the other port is used for interference measurement in the $X$ basis. For $Z$ basis, Bob records the click time of each pulse pair to determine the logic value. For $X$ basis, Bob records which detector clicks.

\textit{(\romannumeral3) Reconciliation}.---We denote the event that only one detector clicks a single click. Bob announces the moment when the single click happens and the corresponding basis. They reserve the events under the same basis and discard the rest. 

\textit{(\romannumeral4) Postprocessing}.---After reconciliation, Alice and Bob perform error correction and privacy amplification to get the final keys. During error correction, the amount of information leaked is denoted as $\rm Leak_{EC}$. After error correction, the keys owned by two users are identical. Alice and Bob perform privacy amplification to extract $l$ bits from raw keys. 

The final key rate considering the finite-key effects is bounded by~\cite{li2024finite}
\begin{equation}
    l = n_z [1-h(\overline{E}_p)] - \rm Leak_{\rm EC} - \log_2\left(\frac{2}{\epsilon_{cor}}\right) - 2\log_2\left(\frac{5}{\epsilon_{sec}}\right),
\end{equation}
where $n_z$ represents the number of detection events in the $Z$ basis. Information leakage during the error correction process is given by ${\rm Leak}_{\rm{EC}} = fn_zh(E_z)$, where $f$ is the error correction efficiency, and $E_z$ denotes the bit error rate in the $Z$ basis. The binary Shannon entropy function $h(x)$ is $-x\log_2x-(1-x)\log_2(1-x)$. The phase error rate in the $Z$ basis, denoted as $\overline{E}_p$, is estimated from the bit error rate in the $X$ basis.
\textcolor{black}{We set the security bound of correctness
$\varepsilon_{\rm cor}=10^{-15}$ and the security bound of secrecy $\epsilon_{\rm sec} = 10^{-10}$~\cite{li2024finite}.} Details about postprocessing and calculations are provided in Appendix.

\section{Experimental demonstration}
\begin{figure*}
  \centering
  \includegraphics[width=0.9\textwidth]{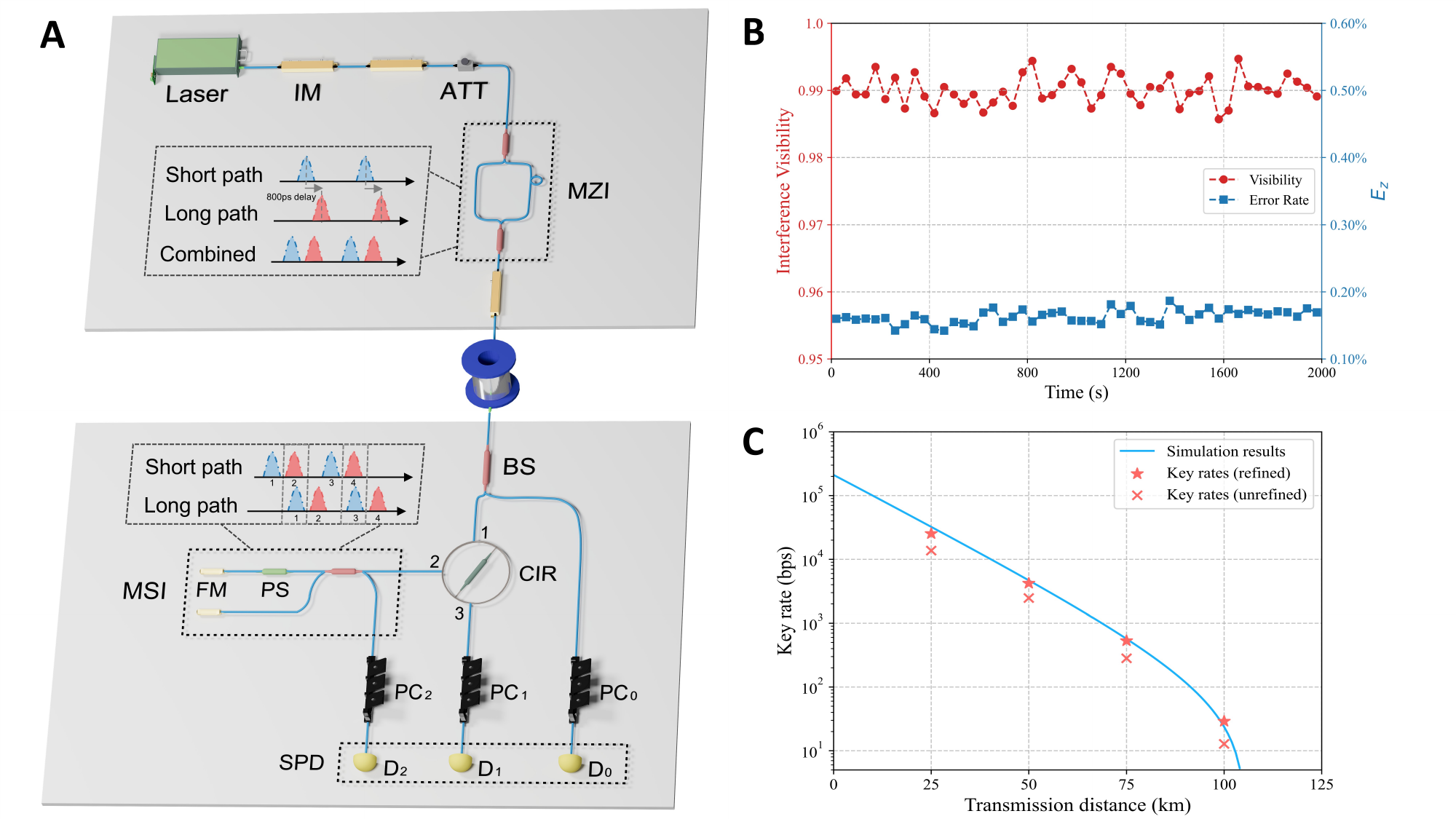}
  \caption{\textbf{Experimental setup and performance of COW-QKD system.} \textbf{a.} Schematic of the experimental setup. IM: intensity modulator; ATT: attenuator; BS: beam splitter; Cir: circulator; FM: Faraday mirror; PS: phase shifter; PC: polarization controller; SPD: superconducting nanowire single-photon detector; MZI: Mach-Zehnder interferometer, consisting of two beam splitters; MSI: Michelson interferometer, consisting of one beam splitter, two Faraday mirrors, and one phase shifter. After passing through MZI, each pulse is split into two subpulses, with a time delay of 800 ps. The arm length differences of these two interferometers are matched. 
  \textbf{b.} Interference visibility and the error rate of $Z$ basis over time. Each data point is derived from detection events accumulated for 40 seconds, with a total of 50 points collected. The interference visibility is calculated from the detection results of the $\ket{\alpha}\ket{\alpha}$ in the $X$ basis. A 100-km optical fiber is inserted during this measurement. The quantum bit error rate is calculated from the detection results of the $\ket{0}\ket{\alpha}$ and $\ket{\alpha}\ket{0}$, prepared in the $Z$ basis without fiber. 
  \textbf{c.} Key rates under different transmission distances. The blue curve represents the simulated key rates based on the actual experimental parameters. The red crosses represent the key rate values directly calculated from the experimental measurements, while the five-pointed stars represent the refined key rates.  \label{fig_cow_exsetup}}
\end{figure*}

We experimentally demonstrate COW-QKD over fiber channels with transmission distances of 25, 50, 75, and 100 km. The attenuation coefficient of the ultra-low-loss optical fibers used in the experiment is less than $0.161$ dB/km. The experimental setup is depicted in Fig.~\ref{fig_cow_exsetup}a.

\textcolor{black}{At Alice's, a narrow linewidth laser (NKT E15) with a linewidth below 0.1 kHz and two intensity modulators (IMs) are used to generate coherent pulses.} After chopping, the  extinction ratio of pulses exceeds 30 dB and the pulse width is less than 200 ps. The repetition rate of this system is 500 MHz. The RF signals and clock synchronization signals required in the experiment are generated by an arbitrary waveform generator (Tabor P9086D).

The optical pulses pass through an asymmetric Mach-Zehnder interferometer (MZI), which consists of two BSs with different arm lengths.
This difference between the two arms splits each input pulse into two separated pulses. The time delay between the two pulses is 800 ps. After MZI, Alice randomly modulates these pulses with an IM according to the logic bit values. Specifically, when Alice chooses $Z$ basis and holds a logic bit value of 0, the first pulse is chopped and the quantum state $\ket{0}\ket{\alpha}$ is prepared. When the logic bit is 1, the prepared quantum state is $\ket{\alpha}\ket{0}$. For the case that $X$ basis is chosen, if the target state is $\ket{0}\ket{0}$, the pulses are chopped before entering the MZI with the first two IMs. If the target state is $\ket{\alpha}\ket{\alpha}$, no modulation is required at the third IM. An attenuator is used to adjust the pulse intensity to the single-photon level.

After transmission through the fiber channel, the optical pulses arrive at Bob's site. A BS with a splitting ratio of $70:30$ is used for passive basis choice. According to the optimization results, 30\% of the input pulses are used for $Z$ basis measurement, while the rest of pulses are used for $X$ basis measurement.
For $X$ basis measurement, we employ an asymmetric Michelson interferometer (MSI) with an arm-length difference matched to MZI at Alice's site, enabling interference between the two pulses in each pair. The MSI consists of a BS and two Faraday mirrors and a phase shifter is integrated into one arm of the interferometer, allowing phase compensation for the relative phase drift. \textcolor{black}{The insertion loss of this phase shifter is relatively low to reduce the effect of loss asymmetries between the two arms of MSI.} All fibers at Bob's site are single-mode, thus maintaining polarization stability. 

In the $X$ basis, the time window is selected to reserve the interference results corresponding to $\ket{\alpha}\ket{\alpha}$, while detection events outside the time window are discarded. As a result, an additional 3 dB loss is introduced. Security analysis requires that the detection efficiencies of the two detectors for $X$ basis measurement must be identical. We randomly remove part of the detection events from detector $\rm D_1$ to ensure that the detection events of the two $X$ basis detectors are approximately equal.
For $Z$ basis measurement, a single-photon detector $\rm D_0$ records the arrival time of Alice's pulses prepared in the $Z$ basis. If the detection event occurs in the former time slot, Bob assigns a logic bit value of 1. If it occurs in the latter time slot, Bob assigns a logic bit value of 0.

After numerical optimization, the probabilities that Alice prepares the states $\ket{0}\ket{0}$, $\ket{\alpha}\ket{\alpha}$, $\ket{0}\ket{\alpha}$, $\ket{\alpha}\ket{0}$ are 10, 10, 40, and 40\%, respectively. The random sequence length used in the implementation is 5000. The detection efficiencies of the three detectors at Bob's side, ${\rm D_0},~{\rm D_1},~{\rm D_2}$, including insertion losses, are approximately $76.2$, $46.0$ and $46.0\%$, respectively (excluding the BS splitting ratio). Their dark count rates are 7, 1, and 11 Hz, respectively. The length of time window is adjusted within the range of 500 to 800 ps. The insertion losses of optical elements in Bob's site and detailed experimental data can be found in Appendix.

\section{Experimental results}
We evaluate the proposed protocol in ultra-low-loss fiber spools with transmission distances of 25, 50, 75, and 100 km. Note that the data size is $10^{11}$ for 25 and 50 km, while it is $10^{12}$ for 75 and 100 km. 
The system performance is presented in Fig.~\ref{fig_cow_exsetup}b.
We measure the interference visibility of $\ket{\alpha}\ket{\alpha}$ in the $X$ basis using experimental parameters similar to those employed in the 100-km implementation.
With 100-km fiber spools, the interference visibility can be kept at approximately 99\% over a period of 2000 seconds.
In addition, we measure the error rate of randomly prepared quantum states, $\ket{0}\ket{\alpha}$ or $\ket{\alpha}\ket{0}$.
Neglecting the effect of dark counts, the bit error rate in the $Z$ basis can be maintained below 0.2\%.

\begin{table*}[htbp]
\centering
\caption{\textbf{Key rates (bps) at different fiber transmission distances.}} 
\label{table_expres}
\setlength{\tabcolsep}{0.3cm}
\renewcommand\arraystretch{1.5}
\begin{tabular}{c@{\hspace{1.2cm}} c@{\hspace{1.2cm}}c@{\hspace{1.2cm}} c@{\hspace{1.2cm}}c}\hline 
Distance &  25 km &  50 km & 75 km & 100 km\\ \hline
Key rate  &  $1.37\times 10^4$  & $2.47\times 10^3$ &  $2.82\times 10^2$ &  12.8                            \\
Key rate (refined) &$2.53\times 10^4$  & $4.21\times 10^3$& $5.31\times 10^2$ &  29.0                        
            \\ 
  \hline
\end{tabular}
\end{table*}

The key rates under different fiber lengths are presented in Fig.~\ref{fig_cow_exsetup}c, with the corresponding values listed in Table~\ref{table_expres}. We observed that the detection counts of $\ket{0}\ket{0}$ in the $X$ basis, $n_{00}^{\rm D_1}$ and $n_{00}^{\rm D_2}$, were relatively high, which may be caused by reflections in MSI. The increase in $n_{00}^{{\rm D}_i}$ ($i\in \{1,2\}$) influences $E_p$, thereby reducing the ultimate key rate. Consequently, we employ two methods to calculate the key rate. The first method directly uses the detection counts in the $X$ basis, $n_{00}^{\rm D_1}$ and $n_{00}^{\rm D_2}$, for key calculation, while the second method infers the values of $n_{00}^{\rm D_1}$ and $n_{00}^{\rm D_2}$ from the detection counts of $\ket{0}\ket{0}$ in the $Z$ basis, $n_{00}^{\rm D_0}$. \textcolor{black}{Noting that the total count rate in the $X$ basis is approximately 70.4\% of that in the $Z$ basis and the detection efficiencies of the two $X$ basis detectors are equal, so we have $n_{00}^{\rm D_1'}=n_{00}^{\rm D_2'}=35.2\% \times n_{00}^{\rm D_0}$.} 
Both methods enable secure key generation over a transmission distance of 100 km.
Employing the second way,
our implementation achieves a key rate of 29 bps at 100 km, while at metropolitan-scale distances ($\sim$ 50 km), the key rate reaches the kilobit-per-second level.

\begin{figure*}
  \centering
  \includegraphics[width=0.7\textwidth]{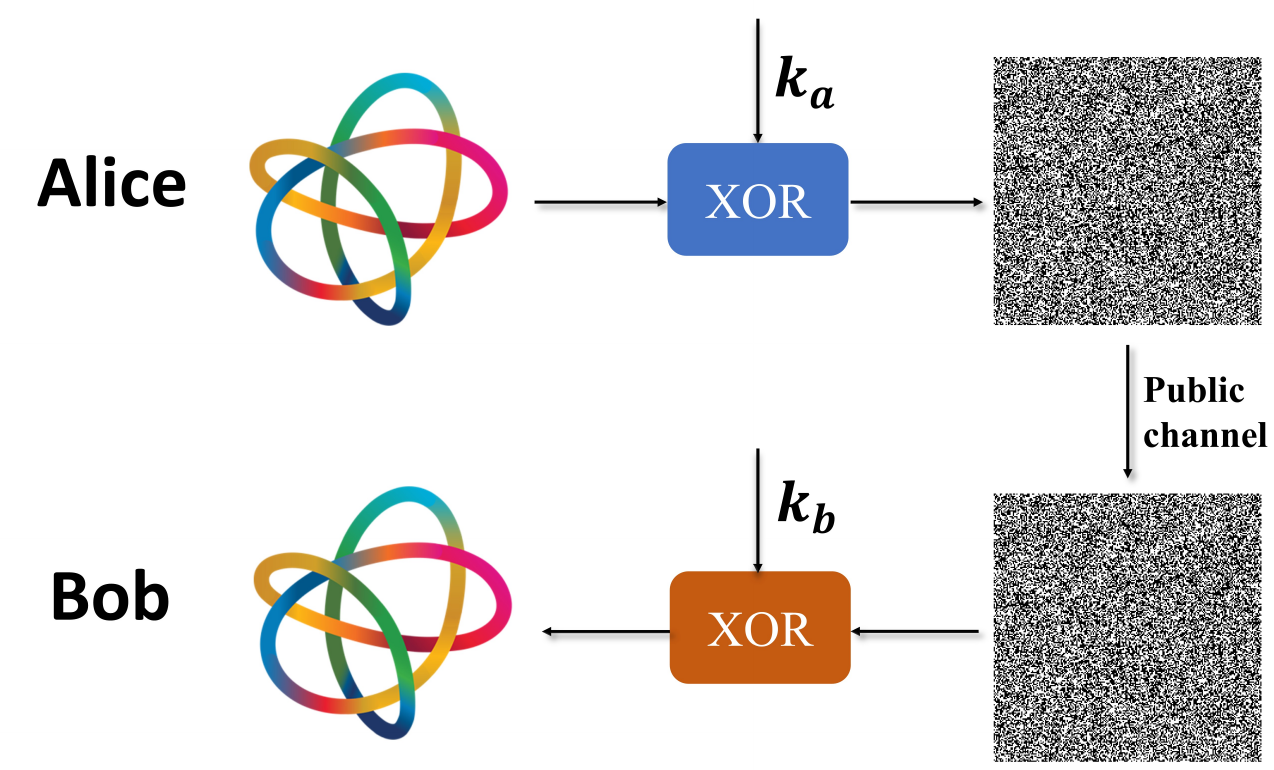}
  \caption{\textbf{Demonstration of the encryption process.} 
    The logo of the International Year of Quantum Science and Technology is used to demonstrate the process of protecting information confidentiality. The logo is first compressed to a size of 6.13 kByte and then converted into a 50,296-bit binary string. Alice performs an XOR operation between her secret key $k_a$ and the binary string to generate the encrypted message, which is transmitted to Bob over a public channel. Upon receiving the string, Bob decrypts it using his corresponding secret key $k_b$, successfully recovering the original binary string and reconstructing the logo.
  \label{fig_cow_encrypt}}
\end{figure*}

To further demonstrate applicability of our scheme, we utilize the secure keys generated over the 100-km fiber link to encrypt and decrypt the logo of the International Year of Quantum Science and Technology, as illustrated in Fig.~\ref{fig_cow_encrypt}. 
During error correction, the number of leaked information is 74,145 and the final key length held by Alice and Bob at 100 km was approximately 58 kb. They encrypted and decrypted the logo independently with their own secret keys. As a result, the logo recovered by Bob was identical to the one originally sent by Alice. This demonstration highlights the ability of the protocol to protect information confidentiality.

\section{DISCUSSION}
In summary, we find that COW-QKD can outperform decoy-state schemes in resisting side-channel attacks on the source and present the experimental implementation of unconditionally secure COW-QKD under finite-key conditions. The transmission distance reaches 100 km, representing the longest reported distance for COW-QKD with information-theoretic security to date.
The scheme preserves the encoding rules of the original protocol, where each logic bit is encoded into a pair of optical pulses and determined by the arrival time of the pulse. Compared to previous schemes, $\ket{0}\ket{0}$ is introduced for security, without adding experimental complexity. We remark that recent studies have demonstrated~\cite{zhang2022experimental,jiang2023side} that since quantum state modulation involves only the vacuum $\ket{0}$ and coherent state $\ket{\alpha}$, it exhibits a potentially substantial advantage over conventional decoy-state protocols~\cite{scarani2009security,xu2020secure} in terms of practical security against source side-channel attacks, with a discussion presented in Appendix.

This work addresses a long-standing open question regarding the feasibility of implementing the COW protocol with practical security. 
The protocol features simple electro-optical modulation and is naturally suitable for high-frequency operation. \textcolor{black}{Employing a pair of arm-length-matched MZI and MSI helps reduce the system repetition rate and maintain phase coherence and stability, thereby improving interference visibility.} With further improvements in system repetition rate, megabit-per-second key rates may be achievable over short distances. \textcolor{black}{Besides, most of the devices involved in this protocol have already been successfully integrated with mature photonic technologies~\cite{li2023high,lin2025integrated,li2025surpassing}. With development of integrated photonics, the system cost and overall performance of COW-QKD are expected to be further improved.}
This work demonstrates that COW-QKD remains of practical value with closing security loopholes, highlighting its potential for future deployment in quantum networks.

\begin{acknowledgments}
This work is supported by the National Natural Science Foundation of China (nos.12522419, U25D8016, and 12274223) and the Fundamental Research Funds for the Central Universities and the Research Funds of Renmin
University of China (no. 24XNKJ14).
\end{acknowledgments}

\appendix
\section{Calculation details}
For the phase error rate $E_p$, the calculation details are presented as follows. First, consider a virtual protocol in which the quantum states prepared in the $Z$ basis are defined as $\ket{0_z} = \ket{0}\ket{\alpha}$ and $\ket{1_z} = \ket{\alpha}\ket{0}$. Consequently, the states in the $X$ basis can be expressed as:
\begin{equation}
\ket{0_x}=\frac{1}{\sqrt{N^+}}(\ket{0_z}+\ket{1_z}), \quad  
\ket{1_x}=\frac{1}{\sqrt{N^-}}(\ket{0_z}-\ket{1_z}),
\end{equation}
where $N^{\pm}=2(1\pm e^{-\mu})$, and $\mu=|\alpha|^2$ is the pulse intensity. 

In this protocol, Alice randomly selects the $Z$ basis and the $X$ basis when preparing quantum states. If $Z$ basis is chosen, Alice prepares $\ket{0_z}$ and $\ket{1_z}$ with equal probability. If $X$ basis is chosen, Alice prepares $\ket{0_x}$ and $\ket{1_x}$ with probabilities $\frac{N^+}{4}$ and $\frac{N^-}{4}$, respectively.

Alice sends the optical pulses to Bob, who uses the same experimental setup as the previous coherent one-way quantum key distribution (COW-QKD) protocol. A beam splitter is used for passive basis choice, and pulses are measured in the corresponding basis. The density matrix remains the same in both the $X$ and $Z$ bases:
\begin{equation}
    \begin{aligned}
        \rho &= \left(\ket{0_z}\bra{0_z}+\ket{1_z}\bra{1_z} \right)/2\\
        &=\left(N^+\ket{0_x}\bra{0_x}+N^-\ket{1_x}\bra{1_x} \right)/4.
    \end{aligned}
    \label{cow_densitym}
\end{equation} 

For the number of clicks from $\ket{k_{x(z)}}$, we have
\begin{equation}
n_{1_x}^{{\rm D}_i}+n_{0_x}^{{\rm D}_i}=n_{1_z}^{{\rm D}_i}+n_{0_z}^{{\rm D}_i}.
\end{equation}
Here, $n_{k_{x(z)}}^{{\rm D}_i}$ represents the number of clicks recorded by Bob’s detector ${\rm D}_i$ ($i\in \{1,2\}$) when Alice sends the state $\ket{k_{x(z)}}$ ($k\in \{0,1\}$).

The bit error rate in the $X$ basis can be expressed as:
\begin{equation}
    \begin{aligned}
        E_x &= \frac{n_{0_x}^{\rm D_2}+n_{1_x}^{\rm D_1}}{n_{0_x}^{\rm D_1}+n_{0_x}^{\rm D_2}+n_{1_x}^{\rm D_1}+n_{1_x}^{\rm D_2}}\\
        &= \frac{n_{0_x}^{\rm D_2}-n_{0_x}^{\rm D_1}+n_{0_z}^{\rm D_1}+n_{1_z}^{\rm D_1}}{n_{0_z}^{\rm D_1}+n_{0_z}^{\rm D_2}+n_{1_z}^{\rm D_1}+n_{1_z}^{\rm D_2}}.
    \end{aligned}
    \label{cow_biterror}
\end{equation}
Since the density matrix remains the same in both the $Z$ and $X$ bases, the phase error rate in COW-QKD protocol equals the bit error rate in the $X$ basis.

In the original COW-QKD protocol, the states $\ket{0_x}$ and $\ket{1_x}$ cannot be directly prepared, making it impossible to directly get $n_{0_x}^{\rm D_1}$ and $n_{0_x}^{\rm D_2}$. Instead, these values can be estimated using the monitoring states $\ket{\alpha}\ket{\alpha}$ and $\ket{0}\ket{0}$, which provide an upper bound $\overline{n}_{0_x}^{\rm D_2}$ and a lower bound $\underline{n}_{0_x}^{\rm D_1}$, respectively. These monitoring states are relatively easy to prepare and their corresponding clicks can be directly recorded. According to Ref.~\cite{zhang2022experimental,jiang2023side,gao2022simple}, the lower bound $\underline{n}_{0_x}^{\rm D_1}$ and upper bound $\overline{n}_{0_x}^{\rm D_2}$ are given by:
\begin{equation}
\begin{aligned}
    \underline{n}_{0_x}^{\rm D_1} &= \frac{N^-}{4N^+}\left(e^{\mu}\frac{n_{\alpha \alpha}^{\rm D_1}}{P_{\alpha\alpha}}+e^{-\mu}\frac{n_{00}^{\rm D_1}}{P_{00}}-2\sqrt{\frac{n_{00}^{\rm D_1}}{P_{00}}\cdot \frac{n_{\alpha \alpha}^{\rm D_1}}{P_{\alpha\alpha}}}\right)\\
    &-\frac{(N^-)^2}{4N^+}\left( e^{\mu} \sqrt{N\frac{n_{\alpha \alpha}^{\rm D_1}}{P_{\alpha\alpha}}}+\sqrt{N\frac{n_{00}^{\rm D_1}}{P_{00}}}\right),
\end{aligned}
    \label{gain_M0}
\end{equation}

\begin{equation}
\begin{aligned}
    \overline{n}_{0_x}^{\rm D_2} &= \frac{1}{4}\left(e^{\frac{\mu}{2}}\sqrt{\frac{n_{\alpha \alpha}^{\rm D_2}}{P_{\alpha\alpha}}}+e^{-\frac{\mu}{2}}\sqrt{\frac{n_{00}^{\rm D_2}}{P_{00}}}\right)^2\\
    &+ \frac{N^-}{4}\left( \frac{e^{\mu}N^-}{4}N+e^{\mu} \sqrt{N\frac{n_{\alpha \alpha}^{\rm D_2}}{P_{\alpha\alpha}}}+\sqrt{N\frac{n_{00}^{\rm D_2}}{P_{00}}}\right).
\end{aligned}
    \label{gain_M1}
\end{equation}
where $n_w^{{\rm D}_i} (w=\alpha\alpha,00)$ represents the number of clicks recorded by ${\rm D}_i$ ($i\in\{1,2\}$) when Alice sends $\ket{\alpha}\ket{\alpha}$ or $\ket{0}\ket{0}$ and $N$ is the total number of sent pulses. $P_{00}$ and $P_{\alpha\alpha}$ represent the probabilities of sending $\ket{0}\ket{0}$ and $\ket{\alpha}\ket{\alpha}$, respectively.

Considering finite-key effects, we use concentration inequalities~\cite{azuma1967weighted,kato2020concentration} to estimate the lower bound of the final key length.
Define $n_1, n_2, \cdots, n_k$ as a set of random variables satisfying $0 \le n_i \le 1$ for $i \in \{1,2,\cdots,k\}$, and define
$\Gamma_i = \sum_{u=1}^i n_u$,
with $f_i$ denoting the $\sigma$-algebra generated by $\{n_1, n_2, \cdots, n_k\}$.
According to Ref.~\cite{li2024finite}, the upper bound of expected value is 
\begin{equation}
    \Gamma_k^* \le \overline{\Gamma}_k^* = \Gamma_k + \Delta_1(a_1,b_1,k,\Gamma_k),
\end{equation}
$\Delta_1(a_1,b_1,k,\Gamma_k) = \left[ b_1+a_1\left(\frac{2\Gamma_k}{k}-1 \right)  \right]\sqrt{k}$.
And the lower bound is 
\begin{equation}
    \Gamma_k^* \ge \underline{\Gamma}_k^* = \Gamma_k - \Delta_2(a_2,b_2,k,\Gamma_k),
\end{equation}
$\Delta_2(a_2,b_2,k,\Gamma_k) = \left[ b_2+a_2\left(\frac{2\Gamma_k}{k}-1 \right)  \right]\sqrt{k}$. The failure probability for estimating the upper bounds is $\varepsilon_a=\exp\left[ \frac{-2(b^2-a^2)}{(1+\frac{4a}{3\sqrt{k}})^2} \right]$ while that for estimating the lower bound is $\varepsilon_a=\exp\left[\frac{-2(b^2-a^2)}{(1-\frac{4a}{3\sqrt{k}})^2} \right]$.

According to Ref.~\cite{curras2021tight}, we can get
\begin{equation}
    \begin{aligned}
        a_1&=a_1(\Gamma_k,k,\varepsilon_a)\\
         &=\frac{3\left( 72\sqrt{k}\Gamma_k(k-\Gamma_k)\ln \varepsilon_a-16k^{3/2}\ln^2\varepsilon_a+9\sqrt{2}(k-2\Gamma_k)C \right)}{4(9k-8\ln \varepsilon_a)(9\Gamma_k(k-\Gamma_k)-2k\ln \varepsilon_a)},
    \end{aligned}
    \label{a1}
\end{equation}
\begin{equation}
    b_1=b_1(a_1,k,\varepsilon_a)=\frac{\sqrt{18a_1^2k-(16a_1^2+24a_1\sqrt{k}+9k)\ln \varepsilon_a}}{3\sqrt{2k}}.
    \label{b1}
\end{equation}

\begin{equation}
    \begin{aligned}
        a_2&=a_2(\Gamma_k,k,\varepsilon_a)\\
         &=-\frac{3\left( 72\sqrt{k}\Gamma_k(k-\Gamma_k)\ln \varepsilon_a-16k^{3/2}\ln^2\varepsilon_a-9\sqrt{2}(k-2\Gamma_k)C \right)}{4(9k-8\ln \varepsilon_a)(9\Gamma_k(k-\Gamma_k)-2k\ln \varepsilon_a)},
    \end{aligned}
    \label{a2}
\end{equation}
\begin{equation}
    b_2=b_2(a_2,k,\varepsilon_a)=\frac{\sqrt{18a_2^2k-(16a_2^2-24a_2\sqrt{k}+9k)\ln \varepsilon_a}}{3\sqrt{2k}},
    \label{b2}
\end{equation}
where $C = \sqrt{-k^2\ln \varepsilon_a(9\Gamma_k(k-\Gamma_k)-2k\ln \varepsilon_a)}$.

The observed values $n_{\alpha\alpha}^{{\rm D}_i}$ and $n_{00}^{{\rm D}_i}$ ($i \in \{1,2\}$) can be directly obtained from the measurement. The total counts $n_{k_z}^{{\rm D}_i}$ ($k \in \{0,1\}$) and $n_z$ for detector $\rm D_0$ in the $Z$ basis can also be obtained experimentally.
Using concentration inequalities, the upper bound of the expected values of these observations is:
\begin{equation}
    n_{w}^{{\rm D}_i*} \le \overline{n}_{w}^{{\rm D}_i*} = n_{w}^{{\rm D}_i} + \Delta_{w}^{{\rm D}_i},
\end{equation}
where $w \in \{00, \alpha\alpha\}$.  

The lower bounds for the expected values of these observations are
\begin{equation}
    n_{w}^{\rm D_1*} \ge \underline{n}_{w}^{\rm D_1*} = n_{w}^{\rm D_1} - \Delta_{w}^{\rm D_1\prime}.
\end{equation}

Consequently, the upper and lower bounds of the gain become:
\begin{equation}
\begin{aligned}
    \underline{n}_{0_x}^{\rm D_1*} &= \frac{N^-}{4N^+}\left(e^{\mu}\frac{\underline{n}_{\alpha \alpha}^{\rm D_1*}}{P_{\alpha\alpha}}+e^{-\mu}\frac{\underline{n}_{00}^{\rm D_1*}}{P_{00}}-2\sqrt{\frac{\overline{n}_{00}^{\rm D_1*}}{P_{00}} \cdot \frac{\overline{n}_{\alpha \alpha}^{\rm D_1*}}{P_{\alpha\alpha}}}\right)\\
    &-\frac{(N^-)^2}{4N^+}\left( e^{\mu} \sqrt{N\frac{\overline{n}_{\alpha \alpha}^{\rm D_1*}}{P_{\alpha\alpha}}}+\sqrt{N\frac{\overline{n}_{00}^{\rm D_1*}}{P_{00}}}\right),
\end{aligned}
    \label{gain_M0_ex}
\end{equation}

\begin{equation}
\begin{aligned}
    \overline{n}_{0_x}^{\rm D_2*} &= \frac{1}{4}\left(e^{\frac{\mu}{2}}\sqrt{\frac{\overline{n}_{\alpha \alpha}^{\rm D_2*}}{P_{\alpha\alpha}}}+e^{-\frac{\mu}{2}}\sqrt{\frac{\overline{n}_{00}^{\rm D_2*}}{P_{00}}}\right)^2\\
    &+ \frac{N^-}{4}\left( \frac{e^{\mu}N^-}{4}N+e^{\mu} \sqrt{N\frac{\overline{n}_{\alpha \alpha}^{\rm D_2*}}{P_{\alpha\alpha}}}+\sqrt{N\frac{\overline{n}_{00}^{\rm D_2*}}{P_{00}}}\right).
\end{aligned}
    \label{gain_M1_ex}
\end{equation}

Finally, the upper bound of the phase error rate is given by:
\begin{equation}
    \overline{E}_p^*=\overline{E}_x^* =\frac{\overline{n}_{0_x}^{\rm D_2*}-\underline{n}_{0_x}^{\rm D_1*}+n_{0_z}^{\rm D_1}+n_{1_z}^{\rm D_1}}{n_{0_z}^{\rm D_1}+n_{0_z}^{\rm D_2}+n_{1_z}^{\rm D_1}+n_{1_z}^{\rm D_2}}.
    \label{cow_ep}
\end{equation}

We also use the concentration inequality again to calculate the upper bound on phase error rate in a case from expected values to observed values. The expected value of  the number of phase errors is $\overline{n}_p^*=N\times\overline{E}_{p}^*$. So the upper bound on the observed value is
\begin{equation}
    n_p \le \overline{n}_p = \overline{n}_p^* + \Delta_p,
    \label{Ep}
\end{equation}
where $\Delta_p=\sqrt{\frac{1}{2}n_{z}\ln\varepsilon_2^{-1}}$ and $\varepsilon_2$ is the failure probability. 
Consequently, we can get 
\begin{equation}
    \overline{E}_p=\overline{n}_p/n_z.
\end{equation}

From the above calculations, we conclude that the parameters required to compute the upper bound of the phase error rate including $n_{\alpha\alpha}^{\rm D_1}, n_{\alpha\alpha}^{\rm D_2}, n_{00}^{\rm D_1}, n_{00}^{\rm D_2}$ can be experimentally measured. The secure key rate computation is thus completed.

\section{Error correction algorithm}
After accumulating enough raw keys, post-processing is performed to extract the final keys. Here, we present the steps of Cascade algorithm used for error correction.

(1) In the first round of error correction, Alice and Bob use the same function  $f_1$ to randomly shuffle the original key while retaining position information. A block length $l$ is chosen, typically set to $l = \frac{K}{E}$, where $K$ is the block length parameter, usually set to 0.73, and  $E$ is the bit error rate. Alice and Bob compute the parity check code for each block, and Alice sends the parity check values to Bob. Bob compares the parity check values. For blocks where the parity check values differ, Bob uses binary search to locate the error bits and flips them. At this point, the number of error bits in all existing blocks is even or zero.

(2) In the $i$ -th round ($ i \geq 2 $), the block length is updated to $\frac{1}{2}$ of the previous round's block length. Alice and Bob use the same shuffle function $f_i$ to randomly shuffle the bits, while recording position information. Alice calculates the parity check value and sends it to Bob, who calculates and compares the values. For blocks where the parity check values differ, Bob uses binary search to locate and flip the error bits. Additionally, Bob backtracks through each round of blocks containing the error, identifying another error bit, and records the position of this error. In subsequent rounds, Bob locates another error in the blocks containing the previously identified error bits, repeating this process. The positions are recorded, and the same operation is repeated in subsequent rounds.

(3) If $i$ is less than the set number of rounds, $i=i+1$ and return to Step 2. Otherwise, the algorithm terminates. In the practical implementation, the frame length is set to 1 Mbit, and the number of error correction rounds is typically set to be less than 4.

\section{Privacy Amplification}
After the error correction, the inconsistent parts in Alice's and Bob's sifted key are corrected, and eventually they obtain exactly the same key.  Privacy amplification needs to be carried out to compress the information that the eavesdropper can obtain to a sufficiently low level. We adopt the privacy amplification method based on Fast Fourier Transform (FFT) technology to optimize the Toeplitz matrix with an algorithm complexity of $O(n\log n)$~\cite{asai2011efficient}. 
For the sifted key of length $n$, it can be multiplied by a Toeplitz matrix of $m \times n$ to obtain a security key of length $m$. The specific steps can be divided into the following four steps.

(1) The first step is to construct a Toeplitz matrix. It is necessary to construct a $m \times n$ random matrix $T$ using $(m+n-1)$ random numbers. Each row of the matrix is obtained by shifting one position to the right of the previous row and adding a new random number.
\begin{equation}
T = \begin{pmatrix}
t_{m} & t_{m+1} & \cdots & t_{m+n-2} & t_{m+n-1} \\
t_{m-1} & t_{m} & \cdots & t_{m+n-3} & t_{m+n-2} \\
\vdots & \vdots & \ddots & \vdots & \vdots \\
t_2 & t_3 & \ddots & t_n & t_{n+1} \\
t_1 & t_2 & \cdots & t_{n-1} & t_n \\
\end{pmatrix}.
\end{equation}

(2) The second step is to expand the $m \times n$ matrix $T$ into a cyclic matrix $T'$ of $(m+n-1) \times (m+n-1)$.
\begin{widetext}
\begin{equation}
T' = \begin{pmatrix}
t_{m} & t_{m+1} & \cdots & t_{m+n-2} & t_{m+n-1} & t_{m-1} & \cdots & t_2 & t_1 \\
t_1 & t_{m} & t_{m+1} & \cdots & t_{m+n-2} & t_{m+n-1} & t_{m-1} & \cdots & t_2 \\
t_2 & t_1 & t_{m} & t_{m+1} & \cdots & t_{m+n-1} & t_{m-1} & \cdots & t_3 \\
\vdots & \ddots & \vdots & \vdots & \ddots & \vdots & \ddots & \ddots & \vdots \\
t_{m-1} & \cdots & t_2 & t_1 & t_{m} & t_{m+1} & \cdots & t_{m+n-2} & t_{m+n-1} \\
t_{m+n-1} &t_{m-1} & \cdots & t_2 & t_1 & t_{m} & t_{m+1} & \cdots & t_{m+n-2} \\
\vdots & \ddots & \ddots & \vdots & \vdots & \ddots & \vdots & \ddots & \vdots \\
t_{m+1} & \cdots & t_{m+n-2} & t_{m+n-1} & t_{m-1} & \cdots & t_2 & t_1 & t_{m} \\
\end{pmatrix}.
\end{equation}
\end{widetext}
Expand the $n$-dimensional original key \textbf{\textit{D}} to the $(m+n-1)$-dimensional column vector \textbf{\textit{D'}} by padding $(m-1)$ zeros.
\begin{equation}
\textbf{\textit{D'}} = \begin{pmatrix}
d_1 \\
d_2 \\
\vdots \\
d_n \\
0 \\
\vdots \\
0 \\
\end{pmatrix}.
\end{equation}

(3) The third step is to calculate the vector \textbf{\textit{B'}} containing the final key by using the FFT and Hadamard product operation.
\begin{equation}
\textbf{\textit{B'}} = T' \textbf{\textit{D'}} = F^{-1} F T' F^{-1} F \textbf{\textit{D}}' = F^{-1}(F T'_{1} .* F \textbf{\textit{D'}}),
\label{B'}
\end{equation}
    where $T'_{1}$ represents the first row of matrix $T'$, $F T'_{1}$ and $F \textbf{\textit{D'}}$ are the FFT operation of $T'_{1}$ and \textbf{\textit{D'}} respectively. After performing Hadamard product operation on $F T'_{1}$ and $F \textbf{\textit{D'}}$, the results need to be subjected to the inverse FFT operation $F^{-1}$ to obtain \textbf{\textit{B'}}.

(4) The fourth step is to perform a modulo-2 operation on each element in vector \textbf{\textit{B'}} and take the first $m$-dimensional vector to obtain the final security key \textbf{\textit{B}}. After postprocessing, Alice and Bob get the final keys.

\section{Side-channel-secure QKD}
In most decoy-state QKD systems~\cite{scarani2009security,xu2020secure}, eavesdroppers may exploit side channels, such as photon frequency spectrum and temporal mode, to extract key information. A recently proposed protocol has demonstrated resistance against source-side external attacks and has been experimentally demonstrated~\cite{zhang2022experimental,jiang2023side}. It is worth noting that the state preparation in our scheme requires only vacuum and non-vacuum states, consistent with side-channel-secure schemes. This feature suggests that COW-QKD may offer advantages over decoy-state schemes in resisting side-channel attacks on the pulses from sources. 

The core idea of side-channel-secure QKD~\cite{zhang2022experimental,jiang2023side} is that a practical source can be regarded as secure if it can be mapped to an ideal virtual source through an appropriate quantum process, so that the key rate with practical sources is equal to that using virtual sources. In the measurement-device-independent protocol, there exist two sources, and we take Alice's source as an example. \textcolor{black}{In the virtual protocol, Alice emits an ideal coherent state $\ket{\alpha_A}$ $(\ket{\alpha_A}=e^{-\mu/2}\ket{0}+\sqrt{1-e^{-\mu}}\ket{\tilde{\alpha}_A})$ together with a vacuum state $\ket{0_A}$. In the real protocol, the vacuum state contains no side-channel space, while all side-channel contributions reside in the non-vacumm part of the coherent state. Therefore, the states emitted by the real source can be expressed in the following form:
\begin{equation}
\begin{aligned}
    &\ket{0}\rightarrow \ket{0},\\
    &\ket{\alpha_A}\rightarrow e^{-\mu/2}\ket{0}+\sqrt{1-e^{-\mu}}\ket{\psi(\tilde{\alpha}_A)},
\end{aligned}
\end{equation}
$\ket{\tilde{\alpha}_A}$ and $\ket{\psi(\tilde{\alpha}_A)}$ have the same probability distribution for different photon-number states when measured in the photon-number space, which means $\fid{0}{\psi(\tilde{\alpha}_A)} = 0$. Consequently, there exists a transformation $U_A$:
\begin{equation}
\begin{aligned}
    &U_A\ket{0_A}= \ket{0_A},\\
    &U_A\ket{\tilde{\alpha}_A}= \ket{\psi(\tilde{\alpha}_A)},
\end{aligned}
\end{equation}}
The same argument applies to Bob’s source. Hence, the two-mode transformation $U_A \otimes U_B$ can map virtual sources to the corresponding real sources. In the ideal case, the side-channel space does not need to be considered. Since the virtual protocol is free from encoding-state side channels, the protocol employing real-life sources preserves the same security.

As the quantum states prepared in the $X$ or $Z$ basis in the side-channel-secure protocol are
\begin{equation}
\begin{aligned}
    \ket{\Psi} &= \frac{1}{\sqrt{2}}\left(\ket{0_z}+\ket{1_z} \right)\\
        &=\left(\sqrt{N^+}\ket{0_x}+\sqrt{N^-}\ket{1_x} \right)/2,
\end{aligned}
\end{equation}
where $\ket{0_z} = \ket{0}_A\ket{\alpha}_B$, $\ket{1_z} = \ket{\alpha}_A\ket{0}_B$, $\ket{0_x}=\frac{1}{\sqrt{N^+}}(\ket{0_z}+\ket{1_z})$, $\ket{1_x}=\frac{1}{\sqrt{N^-}}(\ket{0_z}-\ket{1_z})$ and $N^{\pm}=2(1\pm e^{-\mu})$.
Focusing on the source, COW-QKD is equivalent to this protocol. Building on the security analysis of side-channel-secure protocols, COW-QKD holds potential for resisting source side-channel attacks with experimental simplicity.

\bigskip\noindent\textbf{section S2. Experimental results}\\
\noindent The insertion losses of optical elements in Bob's site are listed in Table~\ref{tab_exp_para} and the experimental results are presented in Table~\ref{exp_res}. In the Michelson interferometer, an optical pulse passes through a beam splitter, reflects off a Faraday mirror, and then passes through the beam splitter again to exit the interferometer, resulting in a total loss of approximately 1.53 dB ($\rm D_1$) and 1.63 dB ($\rm D_2$), respectively.

\begin{table}[ht]
\centering
\renewcommand\arraystretch{1.4}
	\caption{\textbf{The insertion losses of optical elements.}}\label{tab_exp_para}
	\begin{tabular}{@{\hspace{0.75cm}}c@{\hspace{0.75cm}}|@{\hspace{0.75cm}}c@{\hspace{0.75cm}}}\hline \hline
		Optical element &  Insertion loss  \\  \hline
        Cir 1$\rightarrow$2  & 0.51 dB \\
        Cir 2$\rightarrow$3  & 0.47 dB\\ 
		BS-70\%  & 0.2 dB\\
        BS-30\%  & 0.37 dB \\ 
		$\rm{PC_0}$   & 0.1 dB \\
        $\rm{PC_1}$   & 0.08 dB\\
        $\rm{PC_2}$   & 0.17 dB\\
        \hline\hline
	\end{tabular}
\end{table}

\renewcommand\arraystretch{1.7}
\begin{table*}[htbp]
\centering
\caption{\textbf{The experimental data record the measurement results under different fiber lengths.} The recorded data include the intensity $\mu$, total counts in the $Z$ basis $n_{z}$, error rates in the $Z$ basis $E_{z}$, and the counts of the corresponding quantum state at different detectors.} \label{exp_res}
\begin{tabular}{@{\hspace{0.75cm}}c@{\hspace{0.75cm}}|@{\hspace{0.75cm}}c@{\hspace{0.75cm}}|@{\hspace{0.25cm}}cccc}
\hline\hline
\textbf{Fiber length (km)} & $l$ & 25 & 50 & 75 & 100 \\
\hline
\textbf{Intensity} & $\mu$ & $3.50\times 10^{-3}$ & $1.40\times 10^{-3}$ & $5.65\times 10^{-4}$ & $2.43\times 10^{-4}$ \\
\hline
\textbf{Total count} & $n_{z}$ & 25278913 & 4007708 & 6432214 & 1077297 \\
\hline
\textbf{Error rate (\%)} & $E_{z}$ & 0.30 & 0.20 & 0.34 & 0.76 \\
\hline
\multirow{5}{*}{\textbf{$\ket{0}\ket{0}$}} 
& $n_{00}^{\rm D_0}$ & 5409 & 230 & 1572 & 1399 \\
& $n_{00}^{\rm D_1}$ & 22843 & 2194 & 4942 & 840 \\
& $n_{00}^{\rm D_1'}$ & 1904  & 81 & 554 & 493 \\
& $n_{00}^{\rm D_2}$ & 3751 & 566 & 1857 & 1058 \\
& $n_{00}^{\rm D_2'}$ & 1904  & 81 & 554 & 493 \\
\hline
\multirow{2}{*}{\textbf{$\ket{0}\ket{\alpha}$}} 
& $n_{0\alpha}^{\rm D_1}$ & 4668353 & 742102 & 1221403 & 192729 \\
& $n_{0\alpha}^{\rm D_2}$ & 4209875 & 666824 & 1067140 & 172140 \\
\hline
\multirow{2}{*}{\textbf{$\ket{\alpha}\ket{0}$}} 
& $n_{\alpha0}^{\rm D_1}$ & 4202853 & 661568 & 1067269 & 182993 \\
& $n_{\alpha0}^{\rm D_2}$ & 4654407 & 742418 & 1221586 & 204693 \\
\hline
\multirow{2}{*}{\textbf{$\ket{\alpha}\ket{\alpha}$}} 
& $n_{\alpha\alpha}^{\rm D_1}$ & 4413507 & 692141 & 1121394 & 190418 \\
& $n_{\alpha\alpha}^{\rm D_2}$ & 9980 & 1190 & 4228 & 1571 \\
\hline\hline
\end{tabular}
\end{table*}

\end{document}